# CALCULATION OF THE DYNAMIC APERTURE IN THE ANKA STORAGE RING WITH A HIGH-FIELD WAVELENGTH SHIFTER

U. Bandow, D. Einfeld, T. Hezel, B. Krevet, H. O. Moser, R. Rossmanith, J. Schaper
Forschungszentrum Karlsruhe GmbH, Projekt ANKA, Postfach 3640,
D-76021 Karlsruhe, Germany

## 1 ABSTRACT

Third order tracking calculations show that a wavelength shifter on the basis of a commercially available 12 T split-pair solenoid is compatible with the ANKA storage ring.

## 2 INTRODUCTION

Superconducting split-pair solenoids are an attractive and commercially available solution for a high-field wavelength shifter provided that their non-linear field contributions do not spoil the electron optics. We present $3^{rd}$ order tracking calculations concerning the influence of a 12-Tesla split-pair solenoid used as a wavelength shifter in the 2.5 GeV synchrotron light source ANKA [1]. Particles are tracked using MARYLIE [2]. The Lie map representing the non-linear field of the split-pair solenoid is calculated by means of the programs ORBIT/LIEMAP and then read into MARYLIE. The influence on the emittance is calculated by means of DIMAD [3]. The results show that both reduction of the dynamic aperture and emittance blow-up caused by the split-pair solenoid are tolerable if the linear optics is appropriately chosen. In this way, the hard edge of the photon spectrum can be moved significantly towards higher photon energies which widens the range of useful applications in several fields such as tomography, fluorescence or absorption analysis.

## 3 THE SPLIT-PAIR SOLENOID

Fig. 1 shows the coil arrangement of a commercially available 12 Tesla split-pair solenoid [4].

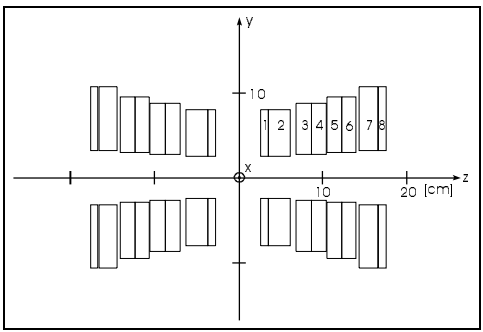

Fig. 1: The arrangement of the eight coils in the cry ostat. z is the direction of the beam.

The magnetic field of each of these coils as calculated with the program TOSCA [5] is shown in fig. 2. The field resulting from their superposition is displayed in fig. 3.

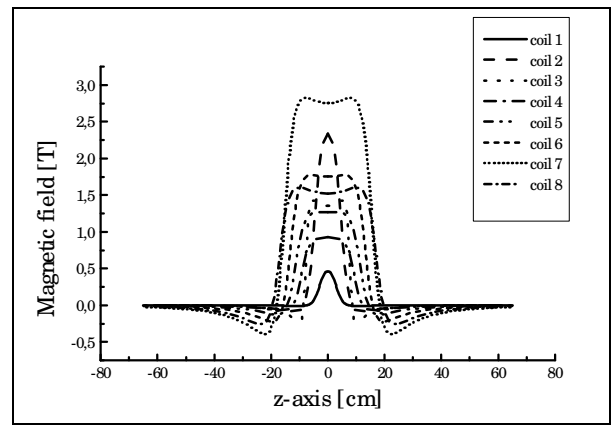

Fig. 2: The field contribution of the individual coils.

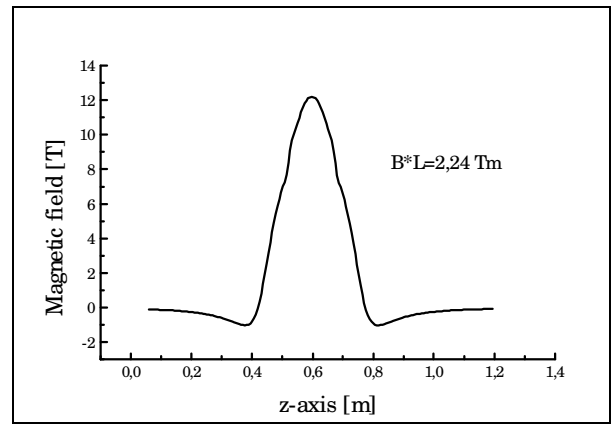

Fig. 3: Superposition of the fields from the individual coils shown in fig. 2.

## 4 INTEGRATION OF THE SPLIT-PAIR SOLENOID MAGNET INTO THE ANKA LATTICE

The integration of the split pair solenoid magnet into the ANKA lattice is shown in fig. 4. The normal 22.5° bends

(B) are followed by a quadrupole triplet (Q$_1$, Q$_2$, Q$_3$) which produces the low horizontal and vertical beta in the insertion region. The split-pair solenoid magnet (SP) deflecting the beam by 15.3° is surrounded by two 7.65° bending magnets (DB).

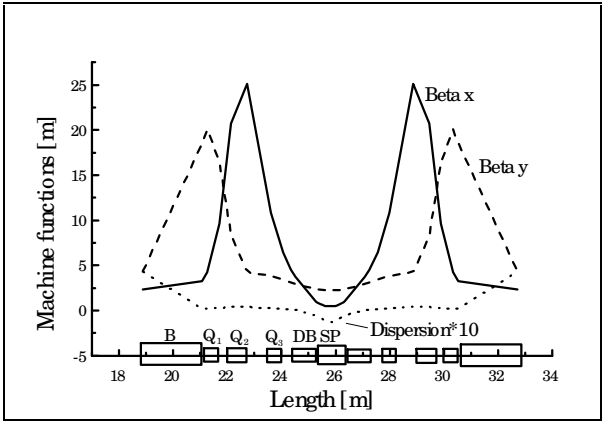

Fig. 4: The optical functions in the insertion device with the split-pair solenoid magnet. Solid line: horizontal beta function. Dashed line: vertical beta function. Dotted line: 10 times horizontal dispersion. B: normal 22.5° bending magnet. Q$_1$, Q$_2$, Q$_3$: quadrupole magnets. DB: 7.65° bending magnet. SP: 15.3° split-pair solenoid magnet.

In order to keep the emittance low the three bending magnets have to be very close to each other in order to minimize the dispersion in the split pair solenoid magnet. In principle, the two 7.65° bending magnets can also be superconducting magnets.

At 2.5 GeV the emittances calculated with DIMAD are 82 nmrad without the split-pair and 87 nmrad with the split-pair and two superconducting 7.65° magnets.

Compared with the original ANKA design the distance between two bends (B) now has a length of 9.47 m (original design 6.47 m). With two sections foreseen for an insertion device opposite to each other the Q values of ANKA are: $Q_X$ = 7.61 and $Q_Y$ = 3.58 (original design values: $Q_X$ = 6.85 and $Q_Y$ = 2.88). The beta values are changed from $\beta_X$ = 18.82 m and $\beta_Y$ = 6.91 m in the original lattice towards $\beta_X$ = 0.49 m and $\beta_Y$ = 2.19 m in the actual lattice.

## 5 CALCULATION OF THE DYNAMIC APERTURE BY SYMPLECTIC TRACKING

The calculation consists of three steps:
1. Calculation of design orbit in the split-pair solenoid: program ORBIT.
2. Generation of a Lie-map for the design orbit: program LIEMAP.
3. Calculation of the dynamic aperture of ANKA by tracking with MARYLIE.

All three programs can be used for general non-linear fields up to 3$^{rd}$ order (and not only for wavelength shifters). The field values have to be provided numerically. ORBIT and LIEMAP were derived from former codes NIN and SCB [6].

ORBIT calculates the design orbit through a non-linear field. An initial reference orbit is assumed. Along this orbit the field and its derivatives are provided by an external magnet code (e.g. TOSCA). The calculation starts with a Runge - Kutta integration and proceeds with an Adams - Moulton predictor-corrector integration.

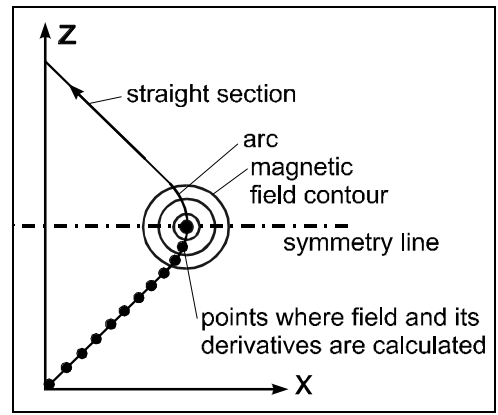

Fig. 5: The auxiliary reference orbit. The magnetic element is positioned in the positive x, z plane. A straight section leads to this element and another straight section marks the end. In a first step it is assumed that the element under consideration has a homogenous field. The trajectory in the magnetic field is approximated by an arc with a constant curvature.

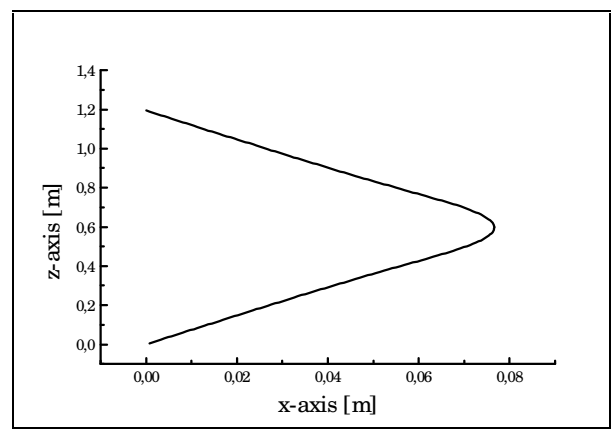

Fig. 6: Optimized design orbit of an electron through the magnetic field of a superconducting split pair solenoid magnet. The transverse momenta at the entrance and exit are almost identical except for their opposite sign.

LIEMAP solves the four differential equations of the design orbit, the matrix and the 3rd and 4th order coefficients of the Lie-map [7].

MARYLIE [2] calculates the dynamic aperture of the storage ring ANKA by symplectic tracking. The map calculated by LIEMAP is read by MARYLIE. The results displayed in fig. 7 show that after introducing the minimum β section the dynamic aperture is only slightly changed by the insertion and is still large compared with the variances of the damped beam ($\Sigma_X$ = 0.25 mm, $\Sigma_Y$ = 0.13 mm).

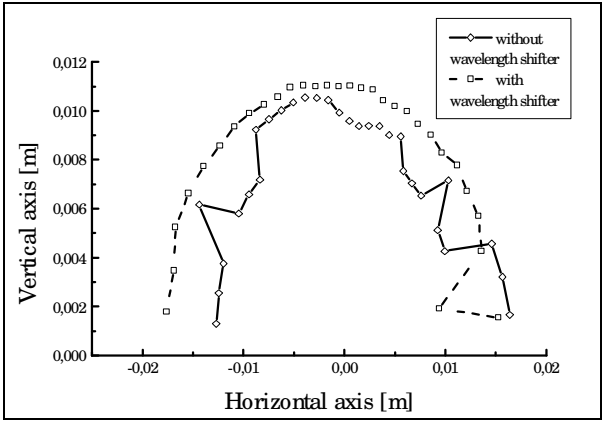

Fig. 7: Dynamic apertures for the low-beta lattice of fig. 4 with and without the wavelength shifter.

## 6 COMPARISON WITH RESULTS FROM BETA

BETA [8] uses a different approach for calculating the dynamic aperture. The split-pair solenoid is simulated by superposing thin multipolar kicks (up to 10th order) on the field of a dipole. These kicks were derived from numerically integrating the magnetic field along different trajectories. Again, the resulting dynamic aperutre is still large enough (fig. 8).

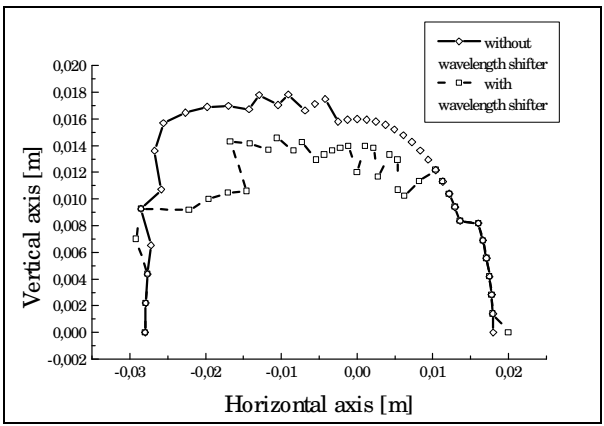

Fig. 8: Dynamic apertures calculated with BETA for the same cases as mentioned in fig. 7.

## 7 THE PHOTON SPECTRUM WITH THE WAVELENGTH SHIFTER

The brilliance from the wavelength shifter and from a standard ANKA bend is shown in fig. 9. The characteristic wavelength is reduced by a factor of 8.

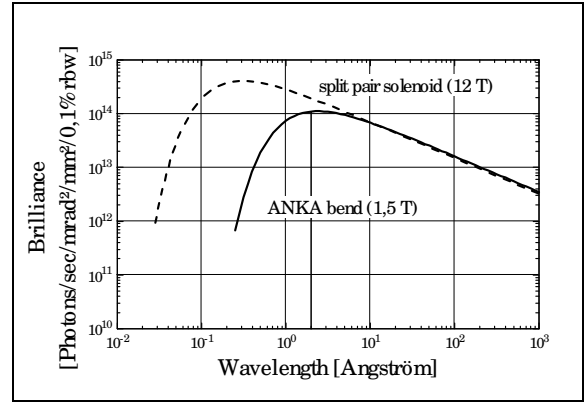

Fig. 9: Brilliance of the wavelength shifter and of a standard ANKA bend.

## 8 SUMMARY

The results show that both reduction of the dynamic aperture and emittance blow-up caused by a commercially available 12 T split-pair magnet as an insertion device in the 2.5 GeV storage ring ANKA are tolerable if the linear optics is appropriately chosen. The wavelength shifter reduces the critical wavelength by a factor of 8 compared to the normal bending magnet of ANKA.

## 9 REFERENCES

[1] H. O. Moser, Industry and synchrotron radiation - prospects using ANKA, Proceedings of this conference, EPAC96, Sitges, Spain

[2] A. J. Dragt et al., MARYLIE 3.0 User's Manual, Department of Physics and Astronomy, University of Maryland, College Park, Maryland 20742, USA

[3] R. V. Servranckx et al., User's Guide to the Program DIMAD, SLAC Report 285 UC-28 (A) May 1985, Stanford Linear Accelerator Center, Stanford, California, USA

[4] Oxford Instruments, Private information

[5] The TOSCA Reference Manual, Vector Fields Limited, 24 Bankside, Kidlington, Oxford, UK

[6] H. O. Moser, A. J. Dragt, Nucl. Instr. and Meth. B24/25 (1987) p. 877

[7] A. J. Dragt and E. Forest, J. Math. Phys. 24 (1983) p. 2734

[8] L. Farvaque, J. L. Laclare and A. Ropert, BETA User's Guide, ESRF-SR/LAT 88-08 (1988), Grenoble, France